\pdfoutput=1

\documentclass[11pt,twoside,a4paper,cmspaper,final,collab]{cms-tdr}
\def\svnVersion{463787P}\def\cmsCernNoTag{CERN-EP-2018-064}\def\cmsCernDate{\today}\def\cmsMessage{Published in Physical Review Letters as \href{http://dx.doi.org/10.1103/PhysRevLett.120.231801}{\doi{10.1103/PhysRevLett.120.231801.}}}

\begin{document}\cmsNoteHeader{HIG-17-035}

\hyphenation{had-ron-i-za-tion}
\hyphenation{cal-or-i-me-ter}
\hyphenation{de-vices}
\RCS$HeadURL: svn+ssh://svn.cern.ch/reps/tdr2/papers/HIG-17-035/trunk/HIG-17-035.tex $
\RCS$Id: HIG-17-035.tex 457789 2018-04-30 17:22:24Z alverson $
\newlength\cmsFigWidth
\ifthenelse{\boolean{cms@external}}{\setlength\cmsFigWidth{0.85\columnwidth}}{\setlength\cmsFigWidth{0.4\textwidth}}
\ifthenelse{\boolean{cms@external}}{\providecommand{\cmsLeft}{upper\xspace}}{\providecommand{\cmsLeft}{left\xspace}}
\ifthenelse{\boolean{cms@external}}{\providecommand{\cmsRight}{lower\xspace}}{\providecommand{\cmsRight}{right\xspace}}
\cmsNoteHeader{HIG-17-035}
\newlength\cmsTabSkip\setlength{\cmsTabSkip}{3ex}

\newcommand{\ytop}{\ensuremath{y_{\PQt}}\xspace}
\newcommand{\sigmod}{\ensuremath{\mu_{\ttbar\PH}}\xspace}
\newcommand{\mf}{\ensuremath{m_\mathrm{f}}\xspace}
\newcommand{\yf}{\ensuremath{y_\mathrm{f}}\xspace}
\newcommand{\hplus}{\ensuremath{\Ph^{+}}\xspace}
\newcommand{\hminus}{\ensuremath{\Ph^{-}}\xspace}
\newcommand{\hplusminus}{\ensuremath{\Ph^{\pm}}\xspace}

\title{Observation of \texorpdfstring{$\ttbar\PH$}{ttbar H} production}

\date{\today}

\abstract{
The observation of Higgs boson production in association
with a top quark-antiquark pair is reported,
based on a combined analysis of proton-proton collision data at
center-of-mass energies of $\sqrt{s}=7$, 8, and 13\TeV,
corresponding to integrated luminosities of up to
5.1, 19.7, and 35.9\fbinv, respectively.
The data were collected with the CMS detector at the CERN LHC.
The results of statistically independent searches for Higgs bosons
produced in conjunction with a top quark-antiquark pair
and decaying to pairs of \PW\ bosons, \cPZ\ bosons, photons, \Pgt\ leptons,
or bottom quark jets are combined to maximize sensitivity.
An excess of events is observed,
with a significance of 5.2 standard deviations,
over the expectation from the background-only hypothesis.
The corresponding expected significance from the standard model
for a Higgs boson mass of 125.09\GeV is 4.2 standard deviations.
The combined best fit signal strength normalized to the standard model prediction
is $1.26{^{+0.31}_{-0.26}}$.
}

\hypersetup{
pdfauthor={CMS Collaboration},
pdftitle={Observation of ttH production},
pdfsubject={CMS},
pdfkeywords={CMS, Higgs boson, ttH}}

\maketitle

Proton-proton ($\Pp\Pp$) collisions at the CERN LHC,
at the center-of-mass (CM) energies of $\sqrt{s}=7$, 8, and 13\TeV,
have allowed direct measurements of the properties of the
Higgs boson~\cite{Aad:1471031,Chatrchyan:1471016,Chatrchyan:1529865}.
In particular,
the 13\TeV data collected so far by the ATLAS~\cite{Aad:2008zzm} and
CMS~\cite{Chatrchyan:2008zzk} experiments have led to
improved constraints on the couplings of the
Higgs boson compared to those performed at the lower
energies~\cite{Aad:2158863},
permitting more precise consistency checks
with the predictions of the standard model (SM) of particle
physics~\cite{SheldonL1961579,PhysRevLett.19.1264,SalamNobel}.
Nonetheless, not all properties of the Higgs boson have
been established,
in part because of insufficiently large data sets.
The lack of statistical precision can be partially overcome by combining
the results of searches in different decay channels of the Higgs boson
and at different CM energies.
Among the properties that are not yet well established
is the tree-level coupling of Higgs bosons to top quarks.

In this Letter, we present a combination of searches for the Higgs boson (\PH)
produced in association with a top quark-antiquark pair (\ttbar),
based on data collected with the CMS detector.
Results from data collected at
$\sqrt{s}=13\TeV$~\cite{Sirunyan:2308650,CMS:2017ttHbbAH,CMS:2017ttHbbSL,CMS:2016ttHgg,Sirunyan:2272260}
are combined with analogous results from
$\sqrt{s}=7$ and 8\TeV~\cite{Khachatryan:1748396}.
As a result of this combination,
we establish the observation of $\ttbar\PH$ production.
This constitutes the first confirmation of the tree-level
coupling of the Higgs boson to top quarks.

A top quark decays almost exclusively to a bottom quark and a \PW\ boson,
with the {\PW} boson subsequently decaying either to a quark and an antiquark
or to a charged lepton and its associated neutrino.
The Higgs boson exhibits a rich spectrum of decay modes that includes
the decay to a
bottom quark-antiquark pair,
a \TT lepton pair,
a photon pair,
and combinations of quarks and leptons
from the decay of intermediate on- or off-shell \PW\ and \cPZ\ bosons.
Thus, $\ttbar\PH$ production gives rise to a wide variety
of final-state event topologies,
which we consider in our analyses and in the combination
of results presented below.

\begin{figure}[btp]
  \centering
    \includegraphics[width=0.2837\textwidth]{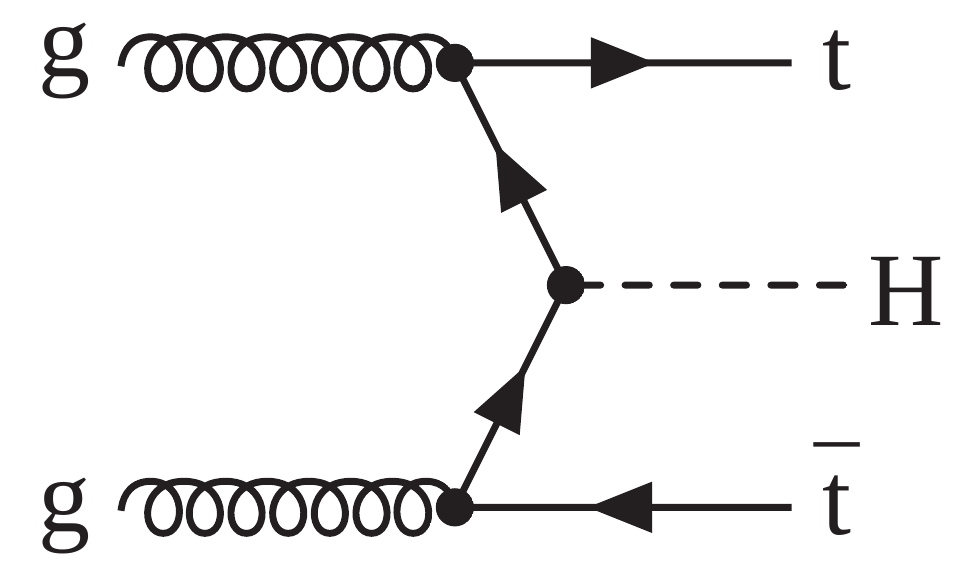}
    \includegraphics[width=0.32\textwidth]{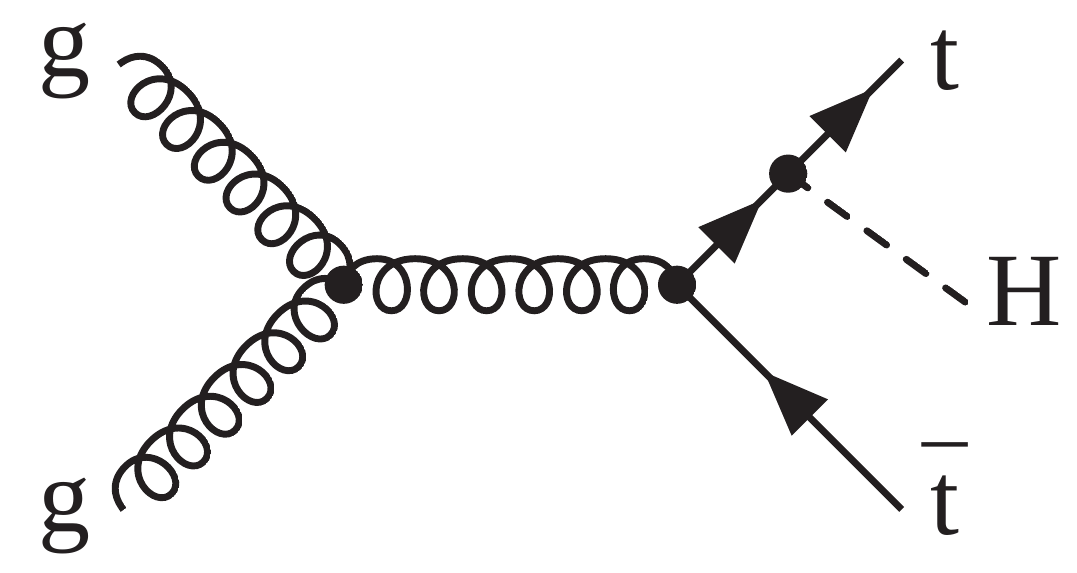}
    \includegraphics[width=0.32\textwidth]{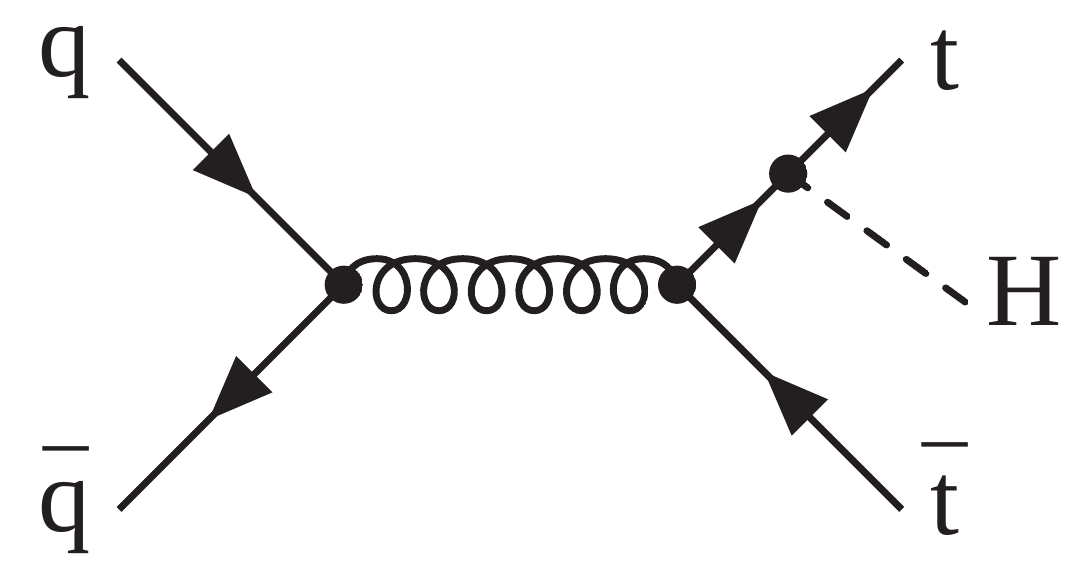}
    \caption{
Example tree-level Feynman diagrams for the $\Pp\Pp\to\ttbar\PH$ production process,
with \cPg\ a gluon, \cPq\ a quark, \cPqt\ a top quark, and \PH a Higgs boson.
For the present study,
we consider Higgs boson decays to a pair of
\PW\ bosons, \cPZ\ bosons, photons, \Pgt\ leptons,
or bottom quark jets.
}
\label{fig:feynman}
\end{figure}

In the SM, the masses of elementary fermions are accounted
for by introducing a minimal set of Yukawa interactions,
compatible with gauge invariance,
between the Higgs and fermion fields.
Following the spontaneous breaking of electroweak
symmetry~\cite{Higgs:1964ia,Englert:1964et,Higgs:1964pj, Guralnik:1964eu, Higgs:1966ev, Kibble:1967sv},
charged fermions of flavor f couple to \PH
with a strength \yf proportional to the mass \mf of those fermions,
namely $\yf = \mf/v$,
where $v\approx 246\GeV$ is the vacuum expectation
value of the Higgs field.
Measurements of the Higgs boson decay rates to down-type fermions
(\Pgt\ leptons and bottom quarks)
agree with the SM predictions within their
uncertainties~\cite{Chatrchyan:Htautau2017,Chatrchyan:Hbb2017}.
However,
the top quark Yukawa coupling (\ytop) cannot be similarly tested
from the measurement of a decay rate
since on-shell top quarks are too heavy
to be produced in Higgs boson decay.
Instead, constraints on \ytop can be obtained
through the measurement of the $\Pp\Pp\to\ttbar\PH$ production process.
Example tree-level Feynman diagrams for this process are shown in Fig.~\ref{fig:feynman}.
To date $\ttbar\PH$ production has eluded definite observation,
although first evidence has been
recently reported by the ATLAS~\cite{Aaboud:2299050}
and CMS~\cite{Sirunyan:2308650} Collaborations.

The overall agreement observed between the SM predictions and data
for the rate of Higgs boson production through gluon-gluon fusion
and for the $\PH\to\gamma\gamma$ decay mode~\cite{Aad:2158863}
suggests that the Higgs boson coupling to top quarks is SM-like,
since the quantum loops in these processes include top quarks.
However, non-SM particles in the loops could introduce terms that
compensate for, and thus mask, other deviations from the SM.
A measurement of the production rate of the tree-level $\ttbar\PH$ process can
provide evidence for,
or against,
such new-physics contributions.

The central feature of the CMS apparatus is a superconducting solenoid
of 6\unit{m} internal diameter,
providing a magnetic field of 3.8\unit{T}.
Within the solenoid volume are a silicon pixel and strip tracker,
a lead tungstate crystal electromagnetic calorimeter,
and a brass and scintillator hadron calorimeter,
each composed of a barrel and two endcap sections.
Forward calorimeters extend the pseudorapidity coverage provided
by the barrel and endcap detectors.
Muons are detected in gas-ionization chambers embedded
in the steel flux-return yoke outside the solenoid.
A detailed description of the CMS detector can be found in
Ref.~\cite{Chatrchyan:2008zzk}.

Events of interest are selected using a
two-tiered trigger system~\cite{Khachatryan:2016bia}
based on custom hardware processors and a farm of commercial processors
running a version of the full reconstruction software optimized for speed.
Offline,
a particle-flow algorithm~\cite{CMS-PRF-14-001} is used to reconstruct and
identify each particle in an event
based on a combination of information
from the various CMS subdetectors.
Additional identification criteria are employed to
improve purities and define the final samples of candidate electrons, muons,
hadronically decaying \Pgt\ leptons
(\tauh)~\cite{Khachatryan:2015dfa,CMS-PAS-TAU-16-002},
and photons.
Jets are reconstructed from particle-flow candidates using
the anti-\kt clustering algorithm~\cite{Cacciari:2008gp}
implemented in the \FASTJET package~\cite{Cacciari:2011ma}.
Multivariate algorithms~\cite{Chatrchyan:2012jua,BTV-16-002}
are used to identify (tag) jets arising from the hadronization
of bottom quarks (\cPqb\ jets)
and discriminate against gluon and light flavor quark jets.
The algorithms utilize observables related to
the long lifetimes of hadrons containing \cPqb\ quarks and
the relatively larger particle multiplicity and mass of \cPqb\ jets
compared to light flavor quark jets.
The \tauh identification is based on the reconstruction
of the hadronic \Pgt\ decay modes
$\Pgt^{-}\to\hminus\PGnGt$, $\hminus\PGpz\PGnGt$, $\hminus\PGpz\PGpz\PGnGt$,
and $\hminus\hplus\hminus\PGnGt$ (plus the charge conjugate reactions),
where \hplusminus\ denotes either a charged pion or kaon.
More details about the reconstruction procedures are given in
Refs.~\cite{Sirunyan:2308650,CMS:2017ttHbbAH,CMS:2017ttHbbSL,CMS:2016ttHgg,Sirunyan:2272260,
Khachatryan:1748396}.

The 13\TeV data employed for the current study were collected in 2016 and
correspond to an integrated luminosity of up to 35.9\fbinv~\cite{CMS-PAS-LUM-17-001}.
The 7 and 8\TeV data,
collected in 2011 and 2012,
correspond to integrated luminosities of up to 5.1
and 19.7\fbinv~\cite{CMS-PAS-LUM-13-001}, respectively.
The 13\TeV analyses are improved relative to the 7 and 8\TeV
studies in that they employ triggers with higher efficiencies,
contain improvements in
the reconstruction and background-rejection methods,
and use more precise theory calculations
to describe the signal and the background processes.
For the 7, 8 and 13\TeV data,
the theoretical calculations of Ref.~\cite{deFlorian:2016spz}
for Higgs boson production
cross sections and branching fractions
are used to normalize the expected signal yields.

The event samples are divided into exclusive categories depending on
the multiplicity and kinematic properties of reconstructed electrons,
muons, \tauh candidates, photons, jets,
and tagged \cPqb\ jets in an event.
Samples of simulated events based on Monte Carlo event generators,
with simulation of the detector response based on
the {\GEANTfour}~\cite{Agostinelli:2002hh} suite of programs,
are used to evaluate the detector acceptance and
optimize the event selection for each category.
In the analysis of data,
the background is,
in general,
evaluated from data control regions.
When this is not feasible,
either because the background process has a very small cross section
or a control region depleted of signal events cannot be identified,
the background is evaluated from simulation
with a systematic uncertainty assigned to account for the known model dependence.
Multivariate
algorithms~\cite{Goodfellow-et-al-2016,Hocker:2007ht,Abazov:2004cs,Abazov:2004ym,Khachatryan:2015ila}
based on deep neural networks,
boosted decision trees,
and matrix element calculations
are used to reduce backgrounds.

At 13\TeV,
we search for $\ttbar\PH$ production in the $\PH\to\bbbar$ decay mode
by selecting events with at least three tagged \cPqb\ jets
and with zero leptons~\cite{CMS:2017ttHbbAH},
one lepton~\cite{CMS:2017ttHbbSL},
or an opposite-sign lepton pair~\cite{CMS:2017ttHbbSL},
where ``lepton'' refers to an electron or muon candidate.
A search for $\ttbar\PH$ production in the $\PH\to\gamma\gamma$
decay mode is performed in events with two reconstructed photons in
combination with reconstructed electrons or muons,
jets, and tagged \cPqb\ jets~\cite{CMS:2016ttHgg}.
The signal yield is  extracted from a fit to the diphoton invariant mass spectrum.
Events with combinations of jets and tagged \cPqb\ jets
and with two same-sign leptons, three leptons, or four leptons
are used to search for $\ttbar\PH$ production in the
$\PH\to\TT$, $\PW\PW^*$, or $\cPZ\cPZ^*$ decay modes~\cite{Sirunyan:2308650,Sirunyan:2272260},
where in this case ``lepton'' refers to an electron, muon, or \tauh candidate
(the asterisk denotes an off-shell particle).
The searches in the different decay channels
are statistically independent from each other.
Analogous searches have been performed with the 7 and 8\TeV
data~\cite{Khachatryan:1748396}.

The presence of a $\ttbar\PH$ signal is assessed by performing a simultaneous
fit to the data from the different decay modes,
and also from the different CM energies as described below.
A detailed description of the statistical methods can be found in
Ref.~\cite{ATLAS:1379837}.
The test statistic $q$ is defined as the negative of
twice the logarithm of the profile likelihood ratio~\cite{ATLAS:1379837}.
Systematic uncertainties are incorporated through the use of nuisance parameters
treated according to the frequentist paradigm.
The ratio between the normalization of the $\ttbar\PH$ production process
and its SM expectation~\cite{deFlorian:2016spz},
defined as the signal strength modifier \sigmod,
is a freely floating parameter in the fit.
The SM expectation is evaluated assuming the
combined ATLAS and CMS value for the mass of the Higgs boson,
which is 125.09\GeV~\cite{Aad:2004386}.
We consider the five Higgs boson decay modes with the
largest expected event yields,
namely $\PH\to\PW\PW^*$, $\cPZ\cPZ^*$, $\gamma\gamma$, \TT, and \bbbar.
Other Higgs boson decay modes and production processes,
including $\Pp\Pp\to\cPqt\PH+\mathrm{X}$ (or $\cPaqt\PH+\mathrm{X}$),
with $\mathrm{X}$ a light flavor quark or \PW\ boson,
are treated as backgrounds and normalized using the predicted SM cross sections,
subject to the corresponding uncertainties.

\begin{figure}[btp]
  \centering
    \includegraphics[width=0.49\textwidth]{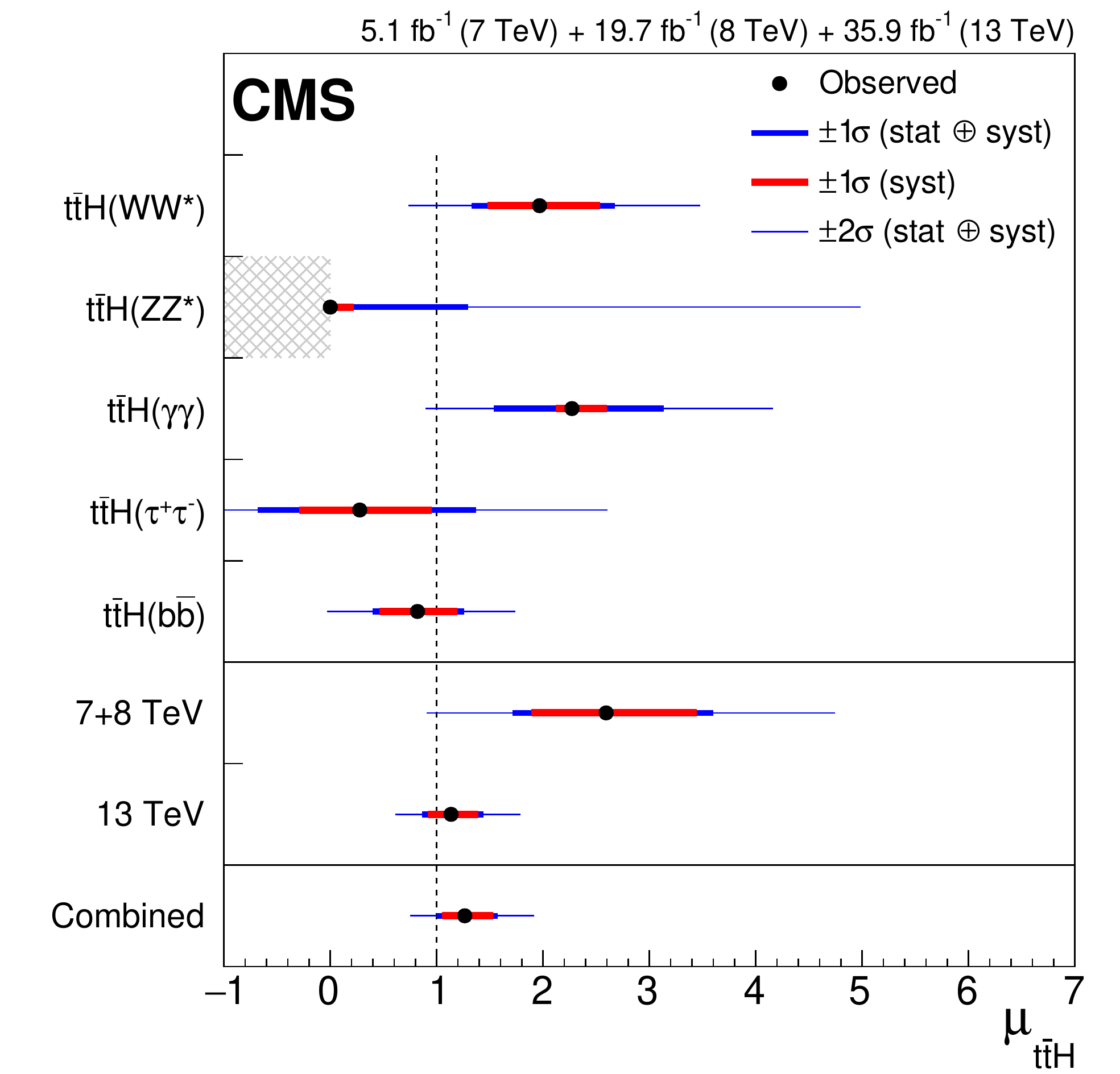}
    \caption{
Best fit value of the $\ttbar\PH$ signal strength modifier \sigmod,
with its 1 and 2 standard deviation confidence intervals ($\sigma$),
for (upper section) the five individual decay channels considered,
(middle section) the combined result for 7+8\TeV alone
and for 13\TeV alone,
and (lower section) the overall combined result.
The Higgs boson mass is taken to be 125.09\GeV.
For the $\PH\to\cPZ\cPZ^*$ decay mode,
\sigmod is constrained to be positive
to prevent the corresponding event yield from becoming negative.
The SM expectation is shown as a dashed vertical line.
}
    \label{fig:summary}
\end{figure}

\begin{table}[htb]
\centering
\topcaption{
Best fit value, with its uncertainty,
of the $\ttbar\PH$ signal strength modifier \sigmod,
for the five individual decay channels considered,
the combined result for 7+8\TeV alone
and for 13\TeV alone,
and the overall combined result.
The total uncertainties are decomposed into their statistical (Stat),
experimental systematic (Expt),
background theory systematic (Thbgd),
and signal theory systematic (Thsig) components.
The numbers in parentheses are those expected for $\sigmod = 1$.}
\label{tab:summary}
\begin{scotch}{@{} l r@{}l c c c c @{}}
           &            &                 & \multicolumn{4}{c}{Uncertainty} \\
Parameter & \multicolumn{2}{c}{Best fit} & Stat & Expt & Thbgd & Thsig \\
\hline\\ [-1.5ex]
\multirow{2}{*}{$\sigmod^{\PW\PW^*}$} & $1.97$ & {}$^{+0.71}_{-0.64}$ &  $^{+0.42}_{-0.41}$ & $^{+0.46}_{-0.42}$ & $^{+0.21}_{-0.21}$ & $^{+0.25}_{-0.12}$  \\[1pt]
    & $\Big($  & {}$^{+0.57}_{-0.54}\Big)$ &  $\Big($$^{+0.39}_{-0.38}$$\Big)$ & $\Big($$^{+0.36}_{-0.34}$$\Big)$ & $\Big($$^{+0.17}_{-0.17}$$\Big)$ & $\Big($$^{+0.12}_{-0.03}$$\Big)$  \\[5pt]
\multirow{2}{*}{$\sigmod^{\cPZ\cPZ^*}$} & $0.00$ & {}$^{+1.30}_{-0.00}$ &  $^{+1.28}_{-0.00}$ & $^{+0.20}_{-0.00}$ & $^{+0.04}_{-0.00}$ & $^{+0.09}_{-0.00}$  \\[1pt]
    & $\Big($  & {}$^{+2.89}_{-0.99}\Big)$ &  $\Big($$^{+2.82}_{-0.99}$$\Big)$ & $\Big($$^{+0.51}_{-0.00}$$\Big)$ & $\Big($$^{+0.15}_{-0.00}$$\Big)$ & $\Big($$^{+0.27}_{-0.00}$$\Big)$  \\[5pt]
\multirow{2}{*}{$\sigmod^{\gamma\gamma}$} & $2.27$ & {}$^{+0.86}_{-0.74}$ &  $^{+0.80}_{-0.72}$ & $^{+0.15}_{-0.09}$ & $^{+0.02}_{-0.01}$ & $^{+0.29}_{-0.13}$  \\[1pt]
    & $\Big($  & {}$^{+0.73}_{-0.64}\Big)$ &  $\Big($$^{+0.71}_{-0.64}$$\Big)$ & $\Big($$^{+0.09}_{-0.04}$$\Big)$ & $\Big($$^{+0.01}_{-0.00}$$\Big)$ & $\Big($$^{+0.13}_{-0.05}$$\Big)$  \\[5pt]
\multirow{2}{*}{$\sigmod^{\TT}$} & $0.28$ & {}$^{+1.09}_{-0.96}$ &  $^{+0.86}_{-0.77}$ & $^{+0.64}_{-0.53}$ & $^{+0.10}_{-0.09}$ & $^{+0.20}_{-0.19}$  \\[1pt]
    & $\Big($  & {}$^{+1.00}_{-0.89}\Big)$ &  $\Big($$^{+0.83}_{-0.76}$$\Big)$ & $\Big($$^{+0.54}_{-0.47}$$\Big)$ & $\Big($$^{+0.09}_{-0.08}$$\Big)$ & $\Big($$^{+0.14}_{-0.01}$$\Big)$  \\[5pt]
\multirow{2}{*}{$\sigmod^{\bbbar}$} & $0.82$ & {}$^{+0.44}_{-0.42}$ &  $^{+0.23}_{-0.23}$ & $^{+0.24}_{-0.23}$ & $^{+0.27}_{-0.27}$ & $^{+0.11}_{-0.03}$  \\[1pt]
    & $\Big($  & {}$^{+0.44}_{-0.42}\Big)$ &  $\Big($$^{+0.23}_{-0.22}$$\Big)$ & $\Big($$^{+0.24}_{-0.23}$$\Big)$ & $\Big($$^{+0.26}_{-0.27}$$\Big)$ & $\Big($$^{+0.11}_{-0.04}$$\Big)$ \\[\cmsTabSkip]
\multirow{2}{*}{$\sigmod^{\text{7+8\TeV}}$} & $2.59$ & {}$^{+1.01}_{-0.88}$ &  $^{+0.54}_{-0.53}$ & $^{+0.53}_{-0.49}$ & $^{+0.55}_{-0.49}$ & $^{+0.37}_{-0.13}$  \\[1pt]
    & $\Big($  & {}$^{+0.87}_{-0.79}\Big)$ &  $\Big($$^{+0.51}_{-0.49}$$\Big)$ & $\Big($$^{+0.48}_{-0.44}$$\Big)$ & $\Big($$^{+0.50}_{-0.44}$$\Big)$ & $\Big($$^{+0.14}_{-0.02}$$\Big)$  \\[5pt]
\multirow{2}{*}{$\sigmod^{13\TeV}$} & $1.14$ & {}$^{+0.31}_{-0.27}$ &  $^{+0.17}_{-0.16}$ & $^{+0.17}_{-0.17}$ & $^{+0.13}_{-0.12}$ & $^{+0.14}_{-0.06}$  \\[1pt]
    & $\Big($  & {}$^{+0.29}_{-0.26}\Big)$ &  $\Big($$^{+0.16}_{-0.16}$$\Big)$ & $\Big($$^{+0.17}_{-0.16}$$\Big)$ & $\Big($$^{+0.13}_{-0.12}$$\Big)$ & $\Big($$^{+0.11}_{-0.05}$$\Big)$  \\[\cmsTabSkip]
\multirow{2}{*}{\sigmod} & $1.26$ & {}$^{+0.31}_{-0.26}$ &  $^{+0.16}_{-0.16}$ & $^{+0.17}_{-0.15}$ & $^{+0.14}_{-0.13}$ & $^{+0.15}_{-0.07}$  \\[1pt]
    & $\Big($  & {}$^{+0.28}_{-0.25}\Big)$ &  $\Big($$^{+0.15}_{-0.15}$$\Big)$ & $\Big($$^{+0.16}_{-0.15}$$\Big)$ & $\Big($$^{+0.13}_{-0.12}$$\Big)$ & $\Big($$^{+0.11}_{-0.05}$$\Big)$  \\[5pt]
\end{scotch}
\end{table}

The measured values of the five independent signal strength modifiers,
corresponding to the five decay channels considered,
are shown in the upper section of Fig.~\ref{fig:summary}
along with their
1 and 2 standard deviation confidence intervals obtained in the
asymptotic approximation~\cite{Khachatryan:1979247}.
Numerical values are given in Table~\ref{tab:summary}.
The individual measurements
are seen to be consistent with each other
within the uncertainties.

We also perform a combined fit,
using a single signal strength modifier \sigmod,
that simultaneously scales the $\ttbar\PH$ production
cross sections of the five decay channels considered,
with all Higgs boson branching fractions fixed to their
SM values~\cite{deFlorian:2016spz}.
Besides the five decay modes considered,
the signal normalizations for the Higgs boson decay modes to gluons,
charm quarks, and $\cPZ\gamma$,
which are subleading and cannot be constrained with existing data,
are scaled by \sigmod.
The results combining the decay modes at 7+8\TeV,
and separately at 13\TeV,
are shown in the middle section of Fig.~\ref{fig:summary}.
The overall result,
combining all decay modes and all CM energies,
is shown in the lower section,
with numerical values given in Table~\ref{tab:summary}.
Table~\ref{tab:summary} includes a breakdown of the total uncertainties
into their statistical and systematic components.
The overall result is
$\sigmod = 1.26 \, {^{+0.31}_{-0.26}}$,
which agrees with the SM expectation
$\sigmod=1$ within 1 standard deviation.

The principal sources of experimental systematic uncertainty
in the overall result for $\sigmod$ stem from the uncertainty
in the lepton and $\cPqb$ jet identification efficiencies and
in the $\tauh$ and jet energy scales.
The background theory systematic uncertainty is dominated by modeling uncertainties
in $\ttbar$ production in association with a $\PW$ boson,
a $\cPZ$ boson,
or a pair of $\cPqb$ or $\cPqc$ quark jets.
The dominant contribution to the signal theory systematic uncertainty
arises from the finite accuracy in the SM prediction
for the $\ttbar\PH$ cross section because of
missing higher order terms
and uncertainties in the proton parton density functions~\cite{deFlorian:2016spz}.

\begin{figure}[btp]
  \centering
    \includegraphics[width=0.49\textwidth]{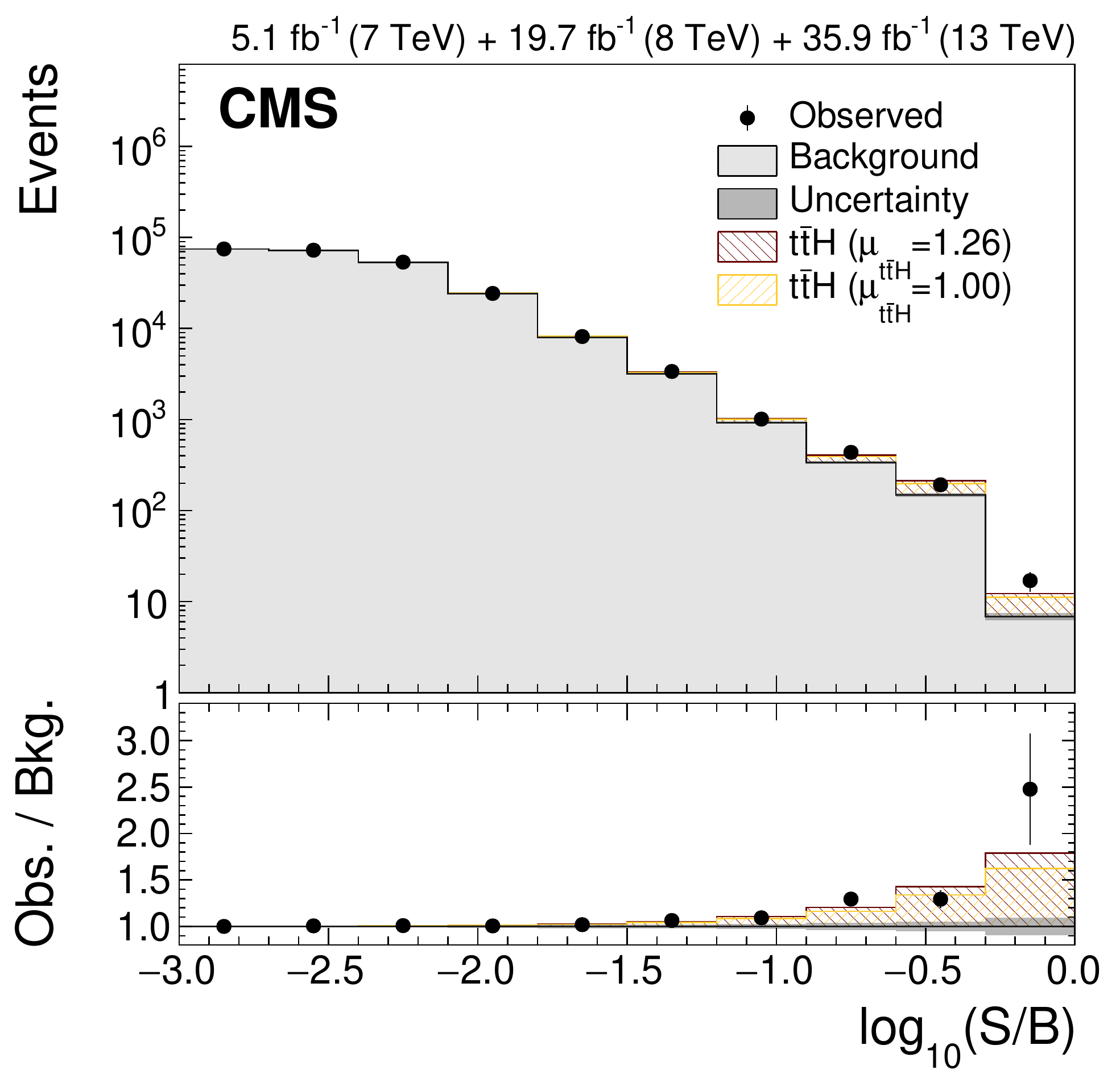}
    \caption{
Distribution of events as a function of the decimal logarithm of $S/B$,
where $S$ and $B$ are the expected post-fit signal (with $\sigmod= 1$)
and background yields, respectively, in each bin of the distributions considered in this combination.
The shaded histogram shows the expected background distribution.
The two hatched histograms,
each stacked on top of the background histogram,
show the signal expectation for the SM
($\sigmod=1$) and the observed ($\sigmod = 1.26$) signal strengths.
The lower panel shows the ratios of the expected signal
and observed results relative to the expected background.
}
    \label{fig:sbplot}
\end{figure}

\begin{figure}[btp]
  \centering
    \includegraphics[width=0.49\textwidth]{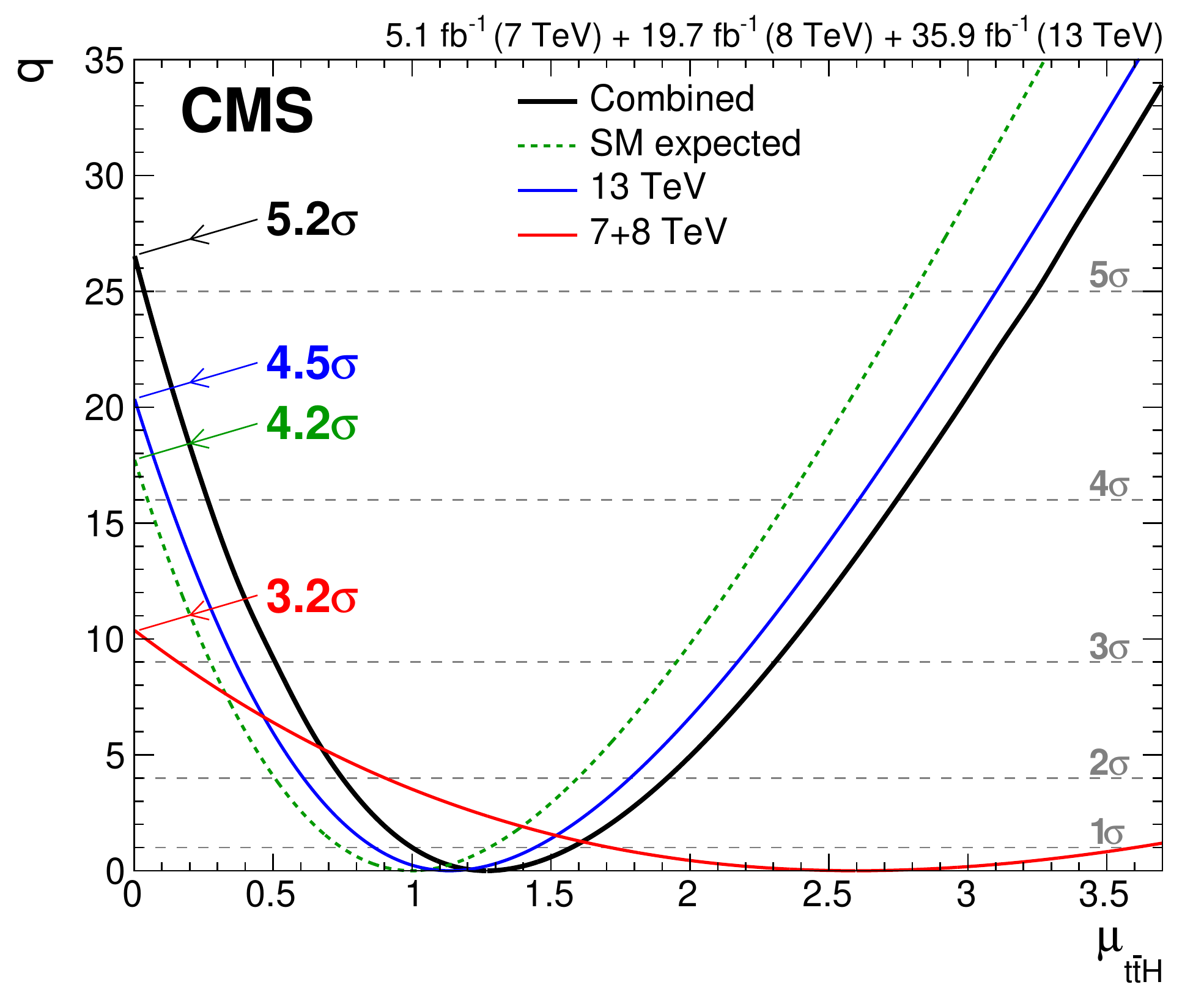}
    \caption{
The test statistic $q$, described in the text,
as a function of \sigmod
for all decay modes at 7+8\TeV and
at 13\TeV, separately,
and for all decay modes at all CM energies.
The expected SM result for the overall combination is also shown.
The horizontal dashed lines indicate the $p$-values for the background-only
hypothesis obtained from the asymptotic distribution of $q$,
expressed in units of the number of standard deviations.
}
    \label{fig:summary2}
\end{figure}

To highlight the excess of data over the expectation from the background-only hypothesis,
we classify each event that enters the combined fit by the ratio $S/B$,
where $S$ and $B$ are the expected post-fit signal (with $\sigmod= 1$)
and background yields, respectively, in each bin of the distributions
considered in the combination.
The distribution of $\log_{10}(S/B)$ is shown in Fig.~\ref{fig:sbplot}.
The main sensitivity at high values of $S/B$ is given by events selected
in the $\PH\to\gamma\gamma$ analysis with a diphoton mass around $125\GeV$
and by events selected in the $\PH\to\TT$, $\PH\to\PW\PW^*$,
and $\PH\to\bbbar$ analyses with high values of the multivariate
discriminating variables used for the signal extraction.
A broad excess of events in the rightmost bins of this distribution is observed,
consistent with the expectation for $\ttbar\PH$ production with a SM-like cross section.

The value of the test statistic $q$ as a function of \sigmod
is shown in Fig.~\ref{fig:summary2},
with \sigmod based on the combination of decay modes
described above for the combined fit.
The results are shown for the combination of all decay modes at 7+8\TeV and
at 13\TeV, separately,
and for all decay modes at all CM energies.
To quantify the significance of the measured $\ttbar\PH$ yield,
we compute the probability of the background-only hypothesis ($p$-value) as
the tail integral of the test statistic using the overall combination
evaluated at $\sigmod=0$ under the asymptotic approximation~\cite{Cowan2011}.
This corresponds to a significance of 5.2 standard deviations
for a one-tailed Gaussian distribution.
The expected significance for a SM Higgs boson with a mass of 125.09\GeV,
evaluated through use of an Asimov data set~\cite{Cowan2011},
is 4.2 standard deviations.

In summary, we have reported the observation of $\ttbar\PH$ production
with a significance of 5.2 standard deviations above the
background-only hypothesis,
at a Higgs boson mass of 125.09\GeV.
The measured production rate is consistent with the standard model
prediction within one standard deviation.
In addition to comprising the first observation of a new Higgs boson
production mechanism,
this measurement establishes the tree-level coupling of the Higgs boson
to the top quark,
and hence to an up-type quark.

\begin{acknowledgments}
We congratulate our colleagues in the CERN accelerator departments for the excellent performance of the LHC and thank the technical and administrative staffs at CERN and at other CMS institutes for their contributions to the success of the CMS effort. In addition, we gratefully acknowledge the computing centers and personnel of the Worldwide LHC Computing Grid for delivering so effectively the computing infrastructure essential to our analyses. Finally, we acknowledge the enduring support for the construction and operation of the LHC and the CMS detector provided by the following funding agencies: BMWFW and FWF (Austria); FNRS and FWO (Belgium); CNPq, CAPES, FAPERJ, and FAPESP (Brazil); MES (Bulgaria); CERN; CAS, MoST, and NSFC (China); COLCIENCIAS (Colombia); MSES and CSF (Croatia); RPF (Cyprus); SENESCYT (Ecuador); MoER, ERC IUT, and ERDF (Estonia); Academy of Finland, MEC, and HIP (Finland); CEA and CNRS/IN2P3 (France); BMBF, DFG, and HGF (Germany); GSRT (Greece); OTKA and NIH (Hungary); DAE and DST (India); IPM (Iran); SFI (Ireland); INFN (Italy); MSIP and NRF (Republic of Korea); LAS (Lithuania); MOE and UM (Malaysia); BUAP, CINVESTAV, CONACYT, LNS, SEP, and UASLP-FAI (Mexico); MBIE (New Zealand); PAEC (Pakistan); MSHE and NSC (Poland); FCT (Portugal); JINR (Dubna); MON, RosAtom, RAS, RFBR and RAEP (Russia); MESTD (Serbia); SEIDI, CPAN, PCTI and FEDER (Spain); Swiss Funding Agencies (Switzerland); MST (Taipei); ThEPCenter, IPST, STAR, and NSTDA (Thailand); TUBITAK and TAEK (Turkey); NASU and SFFR (Ukraine); STFC (United Kingdom); DOE and NSF (USA).
\end{acknowledgments}

\bibliography{auto_generated}

\cleardoublepage \appendix\section{The CMS Collaboration \label{app:collab}}\begin{sloppypar}\hyphenpenalty=5000\widowpenalty=500\clubpenalty=5000\vskip\cmsinstskip
\textbf{Yerevan Physics Institute,  Yerevan,  Armenia}\\*[0pt]
A.M.~Sirunyan,  A.~Tumasyan
\vskip\cmsinstskip
\textbf{Institut f\"{u}r Hochenergiephysik,  Wien,  Austria}\\*[0pt]
W.~Adam,  F.~Ambrogi,  E.~Asilar,  T.~Bergauer,  J.~Brandstetter,  M.~Dragicevic,  J.~Er\"{o},  A.~Escalante Del Valle,  M.~Flechl,  R.~Fr\"{u}hwirth\cmsAuthorMark{1},  V.M.~Ghete,  J.~Hrubec,  M.~Jeitler\cmsAuthorMark{1},  N.~Krammer,  I.~Kr\"{a}tschmer,  D.~Liko,  T.~Madlener,  I.~Mikulec,  N.~Rad,  H.~Rohringer,  J.~Schieck\cmsAuthorMark{1},  R.~Sch\"{o}fbeck,  M.~Spanring,  D.~Spitzbart,  A.~Taurok,  W.~Waltenberger,  J.~Wittmann,  C.-E.~Wulz\cmsAuthorMark{1},  M.~Zarucki
% ...
\vskip\cmsinstskip
% ...
1:  Also at Vienna University of Technology,  Vienna,  Austria\\
% ...
\end{sloppypar}
\end{document}